# Study of Various Dark Matter Halo Profiles in Milky Way and M31 Galaxies within the Standard Cosmology Framework

Darshan Kumar[1], Nisha Rani[2], Deepak Jain[3], Shobhit Mahajan[4] and Amitabha Mukherjee[4]

[1] Institute for Gravitational Wave Astronomy, Henan Academy of Sciences, Zhengzhou 450046, China;
  *kumardarshan@hnas.ac.cn; darshanbeniwal11@gmail.com*

[2] Miranda House, University of Delhi, University Enclave, Delhi 110007, India;
  *nisha.physics@mirandahouse.ac.in*

[3] Deen Dayal Upadhyaya College, University of Delhi, Dwarka, New Delhi 110078, India;
  *djain@ddu.du.ac.in*

[4] Department of Physics and Astrophysics, University of Delhi, Delhi 110007, India;
  *sm@physics.du.ac.in, am@physics.du.ac.in*

**Abstract** In this paper, we study the rotation curves of the Milky Way galaxy (MW) and Andromeda galaxy (M31) by considering their bulge, disk, and halo components. We model the bulge region by the widely accepted de Vaucouleur's law and the disk region by the well established exponential profile. In order to understand the distribution of dark matter in the halo region, we consider three different dark matter profiles in the framework of the standard ΛCDM model namely, Navarro-Frenk-White (NFW), Hernquist and Einasto profiles. We use recent datasets of rotation curves of the Milky Way and Andromeda galaxies. The data consist of rotation velocities of the stars and gas in the galaxy as a function of the radial distance from the center. Using Bayesian statistics, we perform an overall fit including all the components, i.e., bulge, disk and halo with the data. Our results indicate that the NFW and Hernquist profiles are in concordance with the observational data points. However, the Einasto profile poorly explains the behavior of dark matter in both the galaxies.

## 1 INTRODUCTION

A plot of the orbital velocity of stars, gas, or other astronomical objects within a galaxy as they orbit around the center of the galaxy as a function of their distance from the galactic center is referred to as a rotation curve (RC). In a spiral galaxy like our Milky Way or in Andromeda, stars and gas orbit around the galactic center due to the gravitational pull of the combined mass of the matter within the orbit. Following Kepler's laws, one would expect that the orbital velocities of objects within a galaxy should decrease with increasing distance from the galactic center. However, observations reveal otherwise, i.e., the rotation curves of galaxies often exhibit a flat profile at large distances from the galactic center (Rubin & Ford 1970; Rubin et al. 1980; Bosma 1981; Persic & Salucci 1988; Persic et al. 1996; Navarro et al. 1997; Corbelli & Salucci 2000).

This discrepancy between observed rotation curves and the predicted behavior based on the visible mass distribution is one of the key pieces of evidence for the existence of dark matter (DM). Apart from this, indirect evidence like gravitational lensing and galaxy clusters, such as the bullet clusters also points towards the existence of dark matter. However, as yet, no ground based experiment has been able to detect the dark matter particle (Salucci et al. 2021). Hence, for a better understanding of the distribution of dark matter, it becomes crucial to study the rotation curves of the galaxies. It is also important because of the fact that the dynamical evolution of the galaxies and hence of the Universe is affected by the dark matter distribution. Many dark matter profiles have been proposed in the literature. For instance, a phenomenological profile known as the Burkert profile was proposed by Burkert (1995) to describe rotation curves of spiral galaxies (Burkert 1995). The Navarro-Frenk-White (NFW) model that results from $N$-body simulation studies in the Standard $\Lambda$CDM model was proposed by Navarro et al. in 1996 (Navarro et al. 1996). Brownstein (2009) claimed that the rotation curves of low and high surface brightness galaxies can be explained well by the core-modified profile having a constant central density (Brownstein 2009). The characteristic density ($\rho_0$) and the scale length ($r_0$) are the two free parameters in these three profiles. Some more profiles like the Einasto profile (Navarro et al. 2004; Merritt et al. 2006), Generalized NFW (Zhao 1996; An & Zhao 2013) etc., having more than two free parameters have also been proposed in the literature. Comparing the observed rotation curves of the galaxies with these theoretical/phenomenological models can be helpful to infer the distribution of dark matter within galaxies and can further constrain its properties, shedding light on the elusive nature of this mysterious cosmic component.

One of the best choices for this purpose could be our Milky Way galaxy. Being our home galaxy, it is the closest and most accessible galaxy for detailed observations. This proximity allows astronomers to collect high-resolution data and study the dark matter distribution with precision. For details on the measurements of the rotation curve of Milky Way, please refer to Sofue (2017). The observed rotation curves of the Milky Way up to 100 kpc or beyond indicates existence of the dark matter (Sofue 2013; Bhattacharjee et al. 2014; Sofue 2015; Huang et al. 2016). Further, rotation curves of spiral galaxies are essential for studying the mass distribution within galaxies and investigating the properties of their associated dark matter halos. The Andromeda Galaxy (M31), one of the closest spiral galaxies, provides an excellent astrophysical laboratory for studying galaxy formation, evolution, and matter distribution (Rubin & Ford 1970; Carignan et al. 2006; Corbelli et al. 2010). For this reason, we endeavor to explore three dark matter halo profiles, the NFW, Hernquist, and Einasto profiles, in the Milky Way and Andromeda galaxies.

In this work, we use the most updated data of the rotation curve of the MW and M31 galaxy to study the three different dark matter profiles (Sofue 2020, 2015). The paper is organized as follows. In section 2, we discuss the data and methodology. Our results are discussed in section 3. Finally, section 4 is devoted to the discussion and concluding remarks.

## 2 DATA AND METHODOLOGY

In this work, we use data of rotational velocity of two galaxies, Milky Way and M31, as a function of distance from their centre. Data of the Milky Way galaxy are taken from the Tables [5] and [6] of (Sofue 2020). It consists of unified Rotation Curves (RC) from the galactic center to the outer halo at $R \sim 100 \, \mathrm{kpc}$. The data for the M31 galaxy are taken from (Sofue 2015).

Although understanding the real composition of galaxies could be very complicated but in the Newtonian gravity, the square of total rotational velocity can be approximated as the squared sum of three mass components: bulge, disk and halo.

$$v_{\mathrm{tot}}^2(r) = v_b^2(r) + v_d^2(r) + v_h^2(r) \tag{1}$$

where $v_b(r)$, $v_d(r)$ and $v_h(r)$ represent the velocity of the bulge, disk and halo parts as a function of the distance from center of the galaxy respectively. In Newtonian mechanics, the rotation velocity of different parts of the galaxy (namely bulge, disk and halo) can be related to the mass $(M)$ within radius $(r)$ as follows

$$v^2(r) = \frac{GM(r)}{r} \tag{2}$$

$G$ is gravitational constant. From Eq. (2), it is clear that rotational velocity can be determined if the mass distribution of the galaxy is known or vice-versa.

### 2.1 Bulge

The galactic bulge is considered to be spherically symmetric with a de Vaucouleur profile (de Vaucouleurs 1948). Its surface mass density is

$$\Sigma_{br}(r) = \Sigma_b \exp\left[-k\left(\left(\frac{r}{a_b}\right)^{1/4} - 1\right)\right] \tag{3}$$

Here $k = 7.6695$ and $\Sigma_b$ is the surface mass density at the half-mass radius, $(r = a_b)$. The total mass is determined as

$$M_b = 2\pi \int_0^\infty r \, \Sigma_{br}(r) dr = \eta a_b^2 \Sigma_b \tag{4}$$

where $\eta = 22.665$, is a dimensionless constant number.

The mass within a given radius say $r$ can be calculated using

$$M_b(r) = 4\pi \int_0^r \rho_b(r) r^2 dr \tag{5}$$

where $\rho_b(r)$ is the volume density and is calculated using $\Sigma_{br}(r)$ as follows (Binney & Tremaine 2008)

$$\rho_b(r) = -\frac{1}{\pi} \int_r^\infty \frac{d\Sigma_{br}(x)}{dx} \frac{1}{\sqrt{x^2 - r^2}} dx \tag{6}$$

The circular velocity of the bulge can be determined as

$$v_b^2(r) = \frac{4\pi G}{r} \int_0^r \rho_b(r) r^2 dr \tag{7}$$

$\Sigma_b$ and $a_b$ are free parameters for the bulge region.

## 2.2 Disk

The disk of the galaxy is approximated by the exponential profile and its surface mass density can be given as (de Vaucouleurs 1959; Freeman 1970)

$$\Sigma_{dr}(r) = \Sigma_d \exp\left(-\frac{r}{a_d}\right) \tag{8}$$

Here $a_d$ and $\Sigma_d$ are scale radius and central value respectively. The total mass of the exponential disk is

$$M_d = \int_0^\infty 2\pi r \Sigma_{dr} dr = 2\pi \Sigma_d a_d^2 \tag{9}$$

The square of the rotation velocity due to the disk part can be explicitly expressed as (Freeman 1970)

$$v_d^2 = \pi G \Sigma_d \frac{r^2}{a_d} \left[ I_0\left(\frac{r}{2a_d}\right) K_0\left(\frac{r}{2a_d}\right) - I_1\left(\frac{r}{2a_d}\right) K_1\left(\frac{r}{2a_d}\right) \right] \tag{10}$$

where $I_i$ and $K_i$ are the modified Bessel functions.

## 2.3 Dark Halo

The first indirect evidence for the existence of dark matter came from the study of galaxy rotation curves (Rubin et al. 1978). Despite being a fundamental characteristic of galaxies, dark matter remains one of the least understood components. To understand its nature, it is important to explore these rotation curves in detail. Therefore, we focus on analyzing a few prominent dark matter density profiles, namely Navarro–Frenk–White (NFW), Hernquist and Einasto, which are widely used in simulations to model the distribution of dark matter within galaxies.

### 2.3.1 NFW Profile

The Navarro-Frenk-White (NFW) profile is applied in the study of dark matter distribution within galaxies and galaxy clusters. This was proposed by Julio Navarro, Carlos Frenk, and Simon White in 1996. The density profile is written as (Navarro et al. 1996):

$$\rho(r) = \frac{\rho_0}{\left(\frac{r}{h}\right)\left(1 + \frac{r}{h}\right)^2} \tag{11}$$

Here $\rho_0$ and $h$ are scale parameters. From $N$-body cosmological simulation studies in standard $\Lambda$CDM cosmology, it was found that the NFW profile describes the density distribution of dark matter. It suggests that the density of dark matter increases steeply towards the center and gradually decreases in the outer regions. This profile has been found to provide a good fit to the distribution of dark matter in numerical simulations of structure formation in the Universe, offering valuable insights into the large-scale structure and dynamics of the cosmos.

### 2.3.2 Hernquist Profile

This profile was proposed by Lars Hernquist in 1990 and is commonly used to provide analytical expressions for various dynamical quantities namely the gravitational potential, the density of states, the energy distribution function and the surface density. The profile is given as (Hernquist 1990)

$$\rho(r) = \frac{\rho_0}{\left(\frac{r}{h}\right)\left(1 + \frac{r}{h}\right)^3} \tag{12}$$

This profile has been widely employed in both theoretical modeling and observational analyses to study the structural properties of galaxies, their bulges, and the dynamics of their constituent stars and gas. It differs from the NFW profile mainly in the outer parts, where it varies as $r^{-4}$. Also the Hernquist profile has a finite total mass unlike the NFW profile.

### 2.3.3 Einasto Profile

The Einasto profile, proposed by Jaan Einasto in 1965, is a three-dimensional version of the two-dimensional Sérsic (1968) profile (Einasto 1965). The modified functional form of the profile in context of dark matter haloes is

$$\rho(r) = \rho_0 \exp\left[-\left(\frac{r}{h}\right)^{1/n}\right], \tag{13}$$

where $n$ is the Einasto index, a positive number that defines the steepness of the power law. It is clear from Eq. (5) that given $\rho(r)$, one can determine $M(r)$ which can be used to find the velocity of the halo region of the galaxy, i.e.

$$v_h^2(r) = \frac{GM_h}{r} \tag{14}$$

Once the $v_b(r)$, $v_d(r)$ and $v_h(r)$ are calculated using Eqs. (7, 10) and (14), we can use Eq. (1) to get the rotational velocity of the galaxy. We then maximize the likelihood $\mathcal{L} \sim \exp\left(-\chi^2/2\right)$ to put constraints on bulge, disk and dark matter halo parameters. Here chi-square ($\chi^2$) is

$$\chi^2\left(\mathbf{p_b}, \mathbf{p_d}, \mathbf{p_h}\right) = \sum_{i=1}^{N} \frac{\left(v_{\text{tot}}^{\text{th}}\left(r_i; \mathbf{p_b}, \mathbf{p_d}, \mathbf{p_h}\right) - v_{\text{tot}}^{\text{obs}}\left(r_i\right)\right)^2}{\sigma_{v_{\text{tot}}^{\text{obs}}}\left(r_i\right)^2} \tag{15}$$

where $\mathbf{p_b}, \mathbf{p_d}$, and $\mathbf{p_h}$ represent the bulge, disk and halo parameters respectively, e.g. $\mathbf{p_b} \rightarrow \{\Sigma_b, a_b\}$, $\mathbf{p_d} \rightarrow \{\Sigma_d, a_d\}$, and $\mathbf{p_h} \rightarrow \{\rho_0, h, n\}$. In Eq. (15), $v_{\text{tot}}^{\text{obs}}$ and $v_{\text{tot}}^{\text{th}}$ are the observed rotation velocity and total theoretical (contribution of bulge, disk and halo) velocity respectively. The uncertainty in the observed velocity is denoted by $\sigma_{v_{\text{tot}}^{\text{obs}}}$. The total number of data points $(N)$ used for the Milky Way and M31 galaxies are 73 and 55 respectively (Sofue 2020, 2015).

In this analysis, we implement the Markov Chain Monte Carlo (MCMC) analysis using **emcee**, a Python package introduced by Foreman-Mackey and colleagues in 2013 (Foreman-Mackey et al. 2013). After executing the MCMC method to obtain the best fit for all parameters, we determine the confidence levels and their corresponding 68%, 95%, and 99% uncertainties using the **GetDist** Python package (Lewis 2019). For this work, we adopt a wide, uniform prior for all parameters associated with bulge, disk and halo parts.

## 3 RESULTS

In this work, we construct the rotation curves for the Milky Way and M31 galaxies, starting from their central region to the edges of their dark halos. For the inner regions, we adopt the de Vaucouleurs profile for the bulge and the exponential profile for the disk. To model the outer halo regions, we use three different dark matter profiles: NFW, Hernquist, and Einasto.

### 3.1 For Milky Way (MW) galaxy

Using rotational velocity data of the Milky Way galaxy, we analyze the bulge, disk, and dark halo components. By applying Bayesian statistics, we constrain the profile parameters of these components. It is important to note that throughout our analysis, we model the bulge and disk region by de Vaucouleurs and exponential profile respectively while for the dark halo region we consider three different models. The best-fit values of the model parameters are summarized in Table 1.

| S.No. | Model Parameter | NFW | Hernquist | Einasto |
|-------|-----------------|-----|-----------|---------|
| 1. | $a_b$ [kpc] | $1.764^{+0.036}_{-0.038}$ | $1.775^{+0.043}_{-0.035}$ | $1.775^{+0.046}_{-0.035}$ |
| 2. | $\Sigma_b$ [$10^9 \mathrm{M_\odot/kpc^2}$] | $0.595^{+0.014}_{-0.012}$ | $0.592^{+0.013}_{-0.015}$ | $0.593^{+0.013}_{-0.016}$ |
| 3. | $a_d$ [kpc] | $8.817^{+0.224}_{-0.254}$ | $9.237^{+0.205}_{-0.244}$ | $9.130^{+0.304}_{-0.296}$ |
| 4. | $\Sigma_d$ [$10^9 \mathrm{M_\odot/kpc^2}$] | $0.566^{+0.011}_{-0.013}$ | $0.552^{+0.013}_{-0.012}$ | $0.578^{+0.014}_{-0.014}$ |
| 5. | $\rho_0$ [$10^4 \mathrm{M_\odot/kpc^3}$] | $0.712^{+0.546}_{-0.262}$ | $6.309^{+0.495}_{-0.726}$ | $24.866^{+13.721}_{-9.533}$ |
| 6. | $h$ [kpc] | $0.889^{+0.430}_{-0.340} \times 10^3$ | $142.324^{+5.868}_{-15.991}$ | $8.745^{+24.138}_{-5.923}$ |
| 7. | $n$ | - | - | $0.041^{+0.471}_{-0.109}$ |

Table 1: Best fit values of the model parameters of different profiles for the Milky Way galaxy.

From the results, it is clear that the bulge and disk parameters remain consistent across all three dark halo models. However the parameters (bulge and disk) show some correlation among themselves. The parameters $a_b$ and $a_d$ as well as $\Sigma_b$ and $\Sigma_d$ are positively correlated but $a_b$ and $a_d$ show negative correlation with both $\Sigma_b$ and $\Sigma_d$. The pattern remains same in the three profiles, i.e., NFW, Hernquist and Einasto (See Fig. 1, 2, 3 respectively). This suggests that the bulge and disk are independent of the dark halo profile. This can further be seen in the correlation contours shown in the figures. The halo parameters of the dark halo profiles show no correlation with the disk and bulge parameters except in the Einasto profile where $\rho_0$ shows slightly negative correlation with $a_d$. Their stability indicates a clear distinction between the luminous components (bulge and disk) and the dark matter halo in the MW galaxy. However, the dark halo parameters, mainly the scale radius ($h$) and density ($\rho_0$), are not as well-constrained as the bulge and disk parameters. This is because the dark halo parameters are highly correlated with each other.

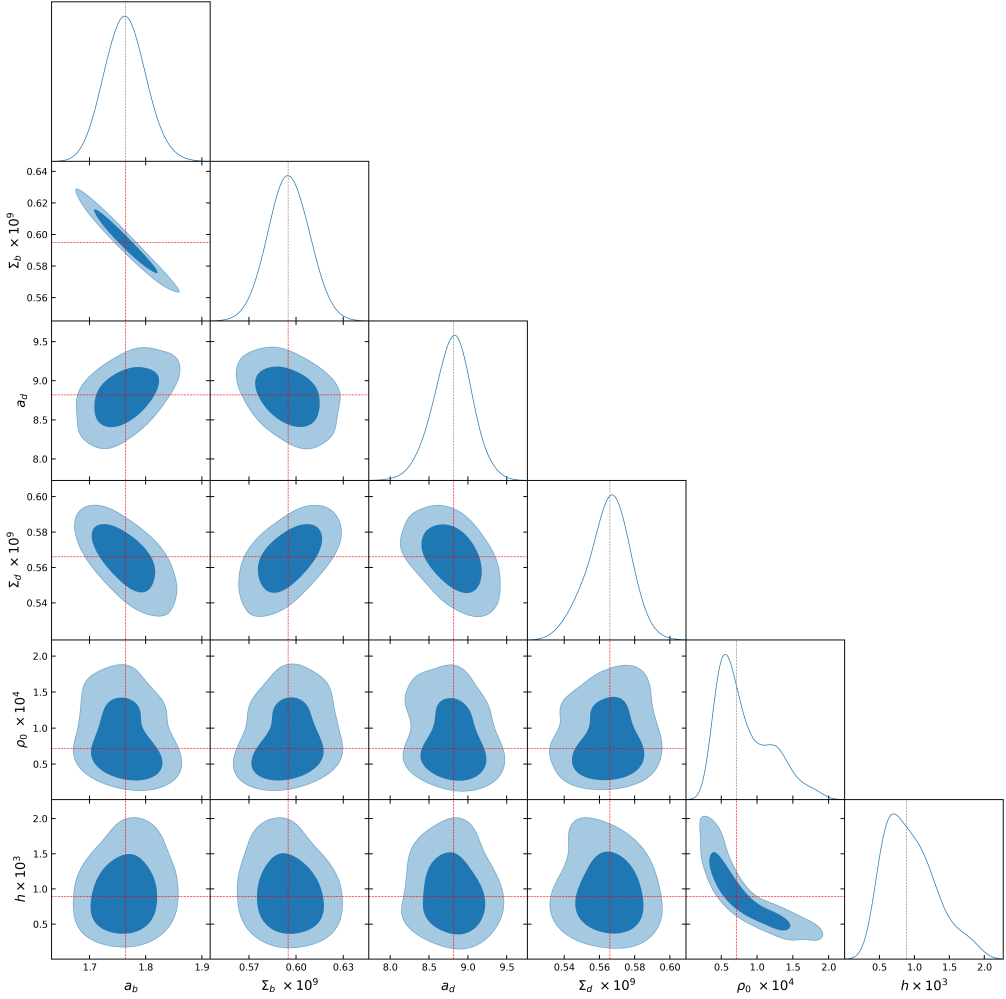

Fig. 1: $1\sigma$ and $2\sigma$ confidence contours and posterior distributions of free parameters of the bulge, disk and halo regions (NFW profile) obtained from the MCMC analysis of the Milky Way galaxy.

The 1D and 2D posterior distributions of the bulge, disk, and halo parameters are illustrated in Figures [1], [2], and [3], corresponding to the NFW, Hernquist, and Einasto halo profiles respectively.

These figures highlight the uncertainties in the parameter space and illustrate the relationships between the different components. The contours indicate strong constraints on the bulge and disk parameters across all three halo profiles. Further, the contours for the dark halo parameters of the NFW profile show a strong negative correlation between the density parameter ($\rho_0$) and the scale radius ($h$) which suggests a degeneracy among them. However for the Hernquist profile, the two parameters do not show any correlation and it remains inconclusive in case of Einasto profile.

### 3.2 For Andromeda (M31) galaxy

We also study the M31 galaxy using its observational rotational velocity data, applying the same dark matter profile as used for the Milky Way. We put constraints on the parameters of these components assuming three different models for the halo region. The obtained best-fit parameters value are listed in Table 2.

The results once again suggest that the bulge and disk parameters remain nearly unchanged across the three halo models, indicating that these luminous components are unaffected by the choice of dark halo profile. The similar behavior can further be observed in the Fig. [4, 5, 6]. These figures show that the disk and bulge parameters follow some correlation with each other but not with the halo parameters, i.e., $a_b$ and $a_d$ as well as $\Sigma_b$ and $\Sigma_d$ are positively correlated while $a_b$ and $\Sigma_d$, $a_b$ and $\Sigma_b$, $a_d$ and $\Sigma_d$ and $a_d$ and $\Sigma_b$ show negative correlations. The contours show no correlation between the halo parameters and bulge-disk parameters. In contrast, the dark halo parameters, particularly the scale radius and density, show weaker constraints. This can be due to the strong correlation between the dark halo parameters.

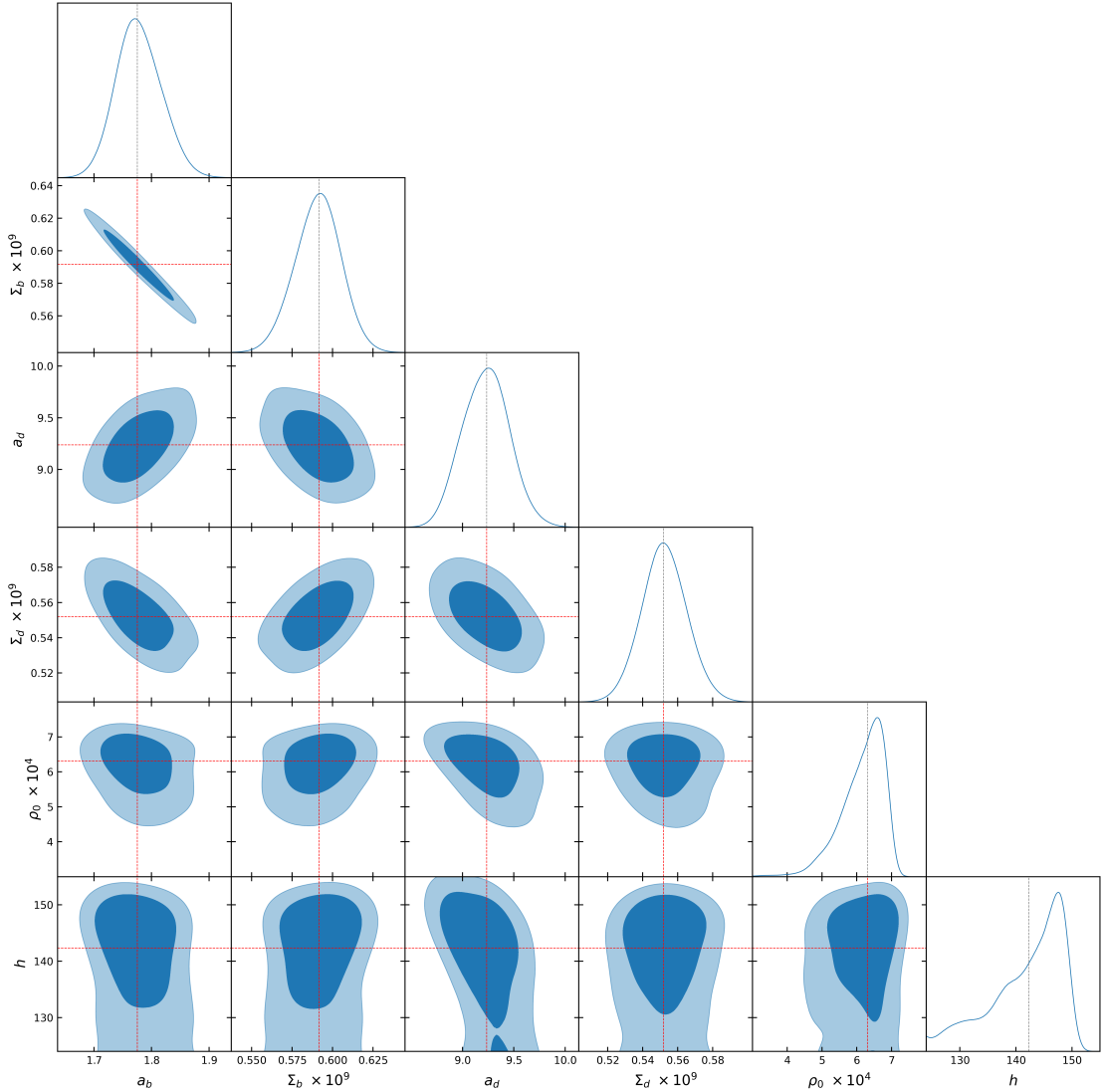

Fig. 2: $1\sigma$ and $2\sigma$ confidence contours and posterior distributions of free parameters of the bulge, disk and halo regions (Hernquist profile) obtained from the MCMC analysis of the Milky Way galaxy.

Figures [4], [5], and [6] show the 1D and 2D posterior distributions for the bulge, disk, and dark halo parameters, corresponding to the NFW, Hernquist, and Einasto profiles for the M31 galaxy.

These distributions provide a clear view of the uncertainties and correlations among the parameters. The contours demonstrate that the bulge and disk parameters are well-constrained across all three halo models. However, the contours for dark halo parameters in M31 also show a pattern similar to the Milky Way, i.e. a strong negative correlation between the density parameter $(\rho_0)$ and the scale radius $(h)$ for the NFW profile but no correlation for the Hernquist profile. Again for the Einasto profile, the correlation remains inconclusive.

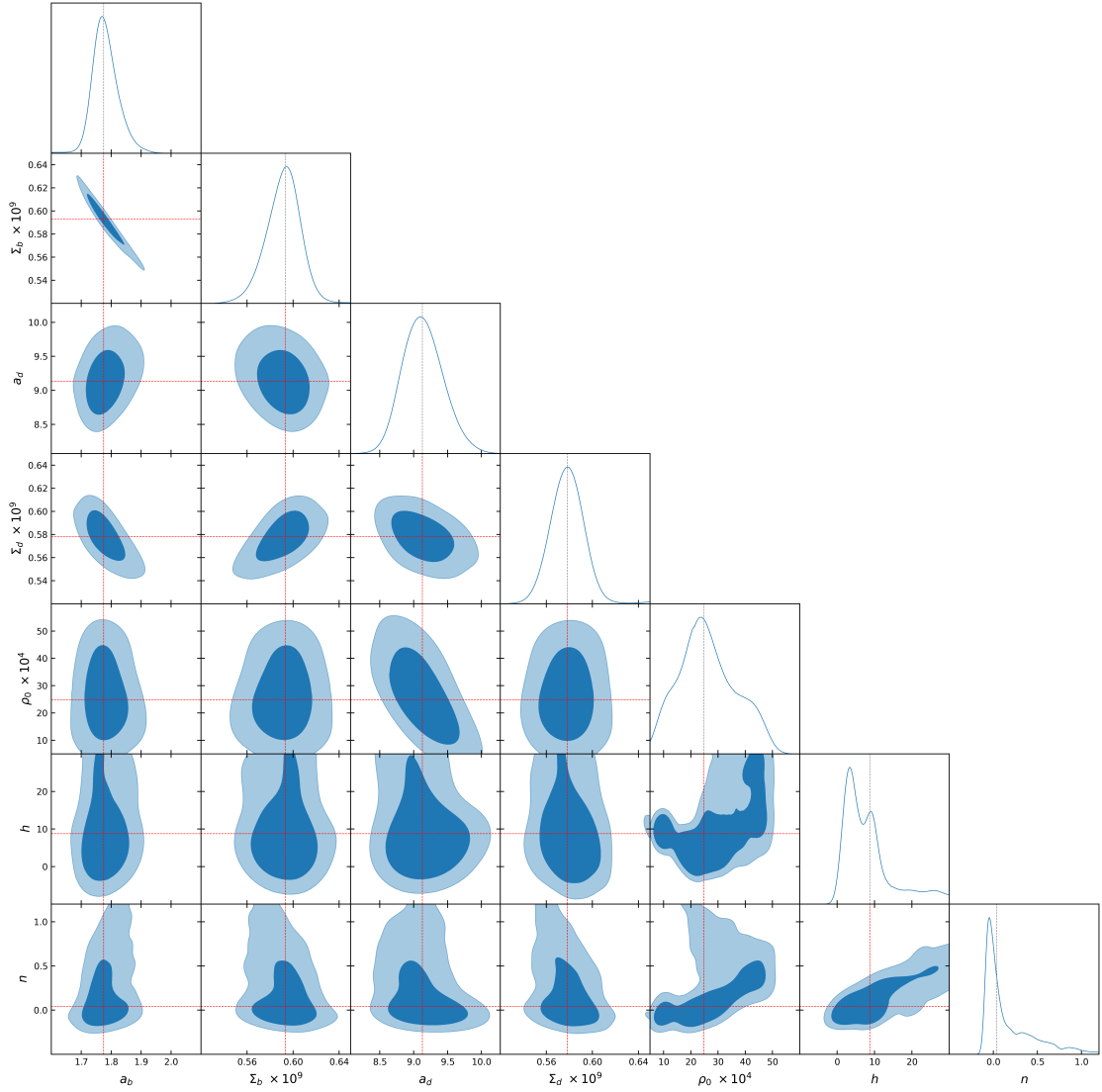

Fig. 3: $1\sigma$ and $2\sigma$ confidence contours and posterior distributions of free parameters of the bulge, disk and halo regions (Einasto profile) obtained from the MCMC analysis of the Milky Way galaxy.

## 4 DISCUSSION AND CONCLUSIONS

In this study, we reconstruct rotation curves of the Milky Way (MW) and Andromeda (M31) galaxies by modeling their bulge, disk, and halo components. For the bulge, we adopt the de Vaucouleurs profile, while the disk is modeled using the established exponential profile. To explore the distribution of dark matter in the halo region, we consider three dark matter profiles within the framework of the standard $\Lambda$CDM model.

 – Navarro-Frenk-White (NFW) profile,
 – Hernquist profile,
 – Einasto profile.

| S.No. | Model Parameters | NFW | Hernquist | Einasto |
|-------|------------------|-----|-----------|---------|
| 1. | $a_b$ [kpc] | $1.280^{+0.117}_{-0.125}$ | $1.252^{+0.174}_{-0.126}$ | $1.309^{+0.324}_{-0.396}$ |
| 2. | $\Sigma_b$ [$10^9 \mathrm{M_\odot/kpc^2}$] | $0.710^{+0.070}_{-0.056}$ | $0.713^{+0.071}_{-0.078}$ | $0.731^{+0.315}_{-0.143}$ |
| 3. | $a_d$ [kpc] | $4.388^{+0.836}_{-0.547}$ | $4.814^{+1.648}_{-1.048}$ | $8.990^{+1.662}_{-0.956}$ |
| 4. | $\Sigma_d$ [$10^9 \mathrm{M_\odot/kpc^2}$] | $0.715^{+0.107}_{-0.120}$ | $0.439^{+0.135}_{-0.301}$ | $0.531^{+0.131}_{-0.144}$ |
| 5. | $\rho_0$ [$10^7 \mathrm{M_\odot/kpc^3}$] | $1.113^{+0.254}_{-0.310}$ | $1.799^{+0.155}_{-0.301}$ | $10.755^{+3.008}_{-0.751}$ |
| 6. | $h$ [kpc] | $16.855^{+2.755}_{-2.089}$ | $18.833^{+0.858}_{-2.084}$ | $4.067^{+0.778}_{-0.712}$ |
| 7. | $n$ | - | - | $0.430^{+0.089}_{-0.131}$ |

Table 2: Best fit values of model parameters of the bulge, disk and halo region considering three different profiles for the halo region of the M31 galaxy.

To constrain the parameters of these profiles, we utilize recent datasets of the MW and M31 galaxies, and apply the Markov Chain Monte Carlo (MCMC) method to maximize the likelihood of the models considered. Although similar analyses have been done earlier (Boshkayev et al. 2024; Lin & Li 2019; Ou et al. 2024), we believe that our work has some new aspects.

1. In earlier studies, Boshkayev et. al reproduced rotation curves of the M31 galaxy considering six different phenomenological halo profiles including NFW using 2015 data (Boshkayev et al. 2024). In contrast, our analysis of the M31 and MW galaxy uses the more recent (2020) data, providing updated constraints on the dark matter distribution. We have also considered two different models for the dark halo, i.e. Hernquist and Einasto which are not considered in their work. Our results show that the bulge-disk parameters are independent on the choice of the dark matter profile which are consistent with their results.

2. Lin & Li (2019) investigated the dark matter profile of the Milky Way using the observed rotation curve data out to 100 kpc. For this, they considered four different dark matter profiles, i.e. the Burkert, the core-modified, the pseudo-isothermal and the NFW profile. They considered seven different models for the bulge part and four for the disk. The authors also took into account the gaseous contribution by considering two gaseous models. With different permutations and combinations, they concluded that the NFW profile fits the data better than the other profiles considered in their work (Lin & Li 2019). We also find that NFW profile matches well with the observational data of rotation curves of the galaxy. In 2023, X. Ou et. al also reconstructed circular velocity curve for the Milky Way for $R \sim 6 - 27.5$ kpc. Further, they studied the dark halo density using the Einasto and Generalized NFW (gNFW) profile and claimed that the Einasto profile fits better than the gNFW (Ou et al. 2024). They further discussed that the Einasto DM halo model predicts a circular velocity significantly lower than the values computed from the enclosed mass estimates which is something similar to what we are obtaining. However in contrast to the X. Ou results, the rotation curves of the MW galaxy using NFW and Hernquist matches better with the observational data as compared to the Einasto Profile in our work. This may be because of the simplicity, computational efficiency, and cosmological connections of the NFW and Hernquist profile while on the other hand the Einasto profile, though more flexible due to an additional parameter, could be often unnecessarily complex and computationally intensive for typical observational data. It is also important to note that the Hernquist DM profile used here has not been explored along with the recent observed data, to the best of our knowledge. We repeat the whole analysis for the M31 galaxy which we believe is extension of the work done in the past.

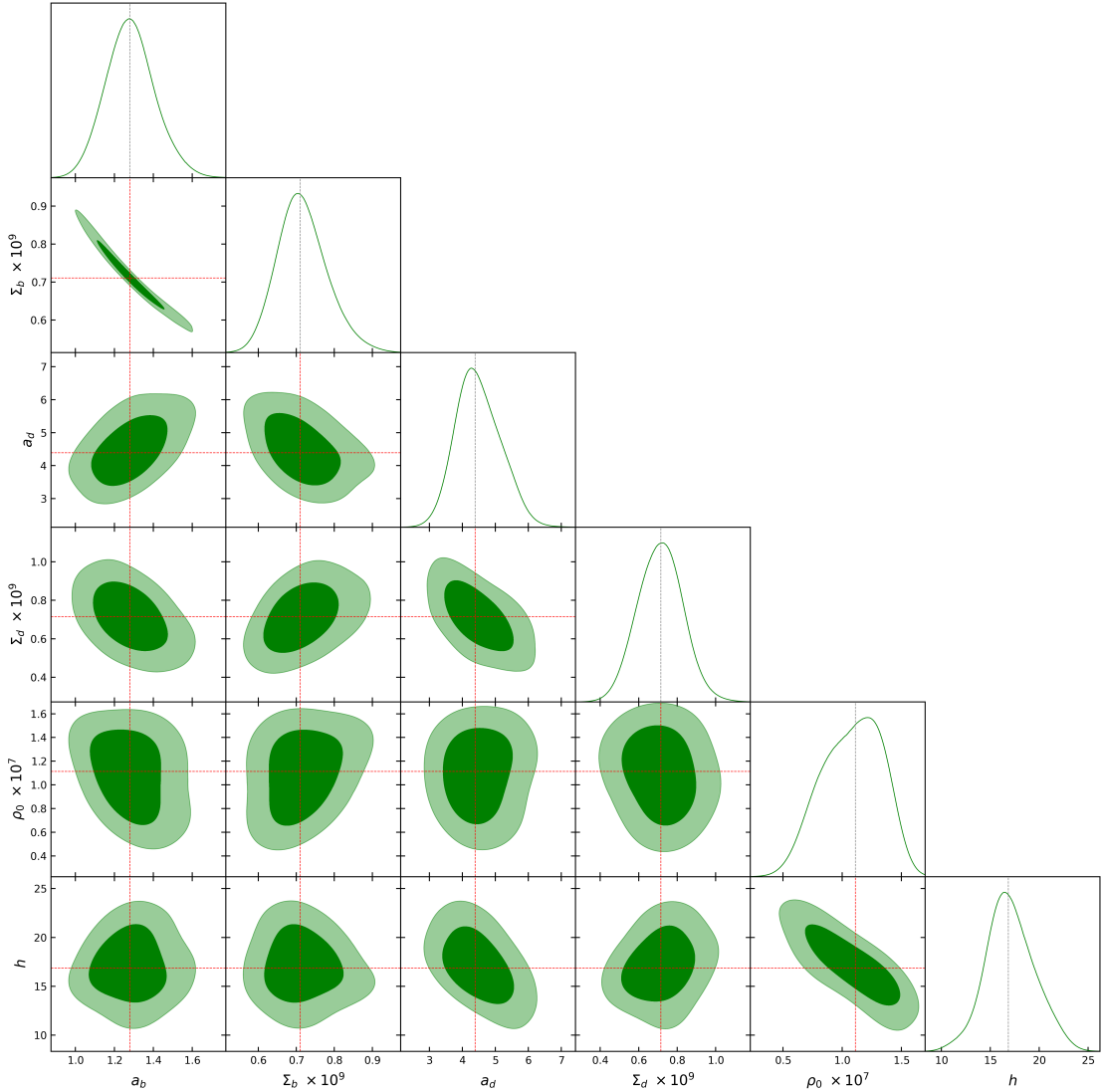

Fig. 4: $1\sigma$ and $2\sigma$ confidence contours and posterior distributions of free parameters of the bulge, disk and halo regions (NFW profile) obtained from the MCMC analysis of the M31 galaxy.

3. De-Chang Dai et. al analyzed the existing rotation-curve data of the Milky Way and M31 galaxies upto large distances under two different cosmological models, i.e., $\Lambda$CDM and MOND (Modified Newtonian Dynamics). Their findings show that a systematic downward trend in the RC of the two galaxies is similar to the one obtained in $\Lambda$CDM EAGLE simulation (Dai et al. 2022). Further, the work by R. D'Agostino et. al employed recent MW data to study the NFW profile in both standard and modified (Yukawa) cosmology (D'Agostino et al. 2024). In both of these works, authors studied RC of the galaxies in different cosmological scenarios and concluded that the rotation curves obtained in the $\Lambda$CDM model resonate better with the observations. Hence, we decided to explore three different dark matter profiles of the two galaxies in the standard cosmological regime. The best fit values of the free parameters that we obtained for both the galaxies are in concordance with the values obtained by R. D'Agostino et. al under the $\Lambda$CDM model. Our work goes beyond the earlier work by incorporating more than just the NFW profile. We also include the Hernquist and Einasto profiles. These models are applied to both the MW and M31 galaxies using the latest datasets.

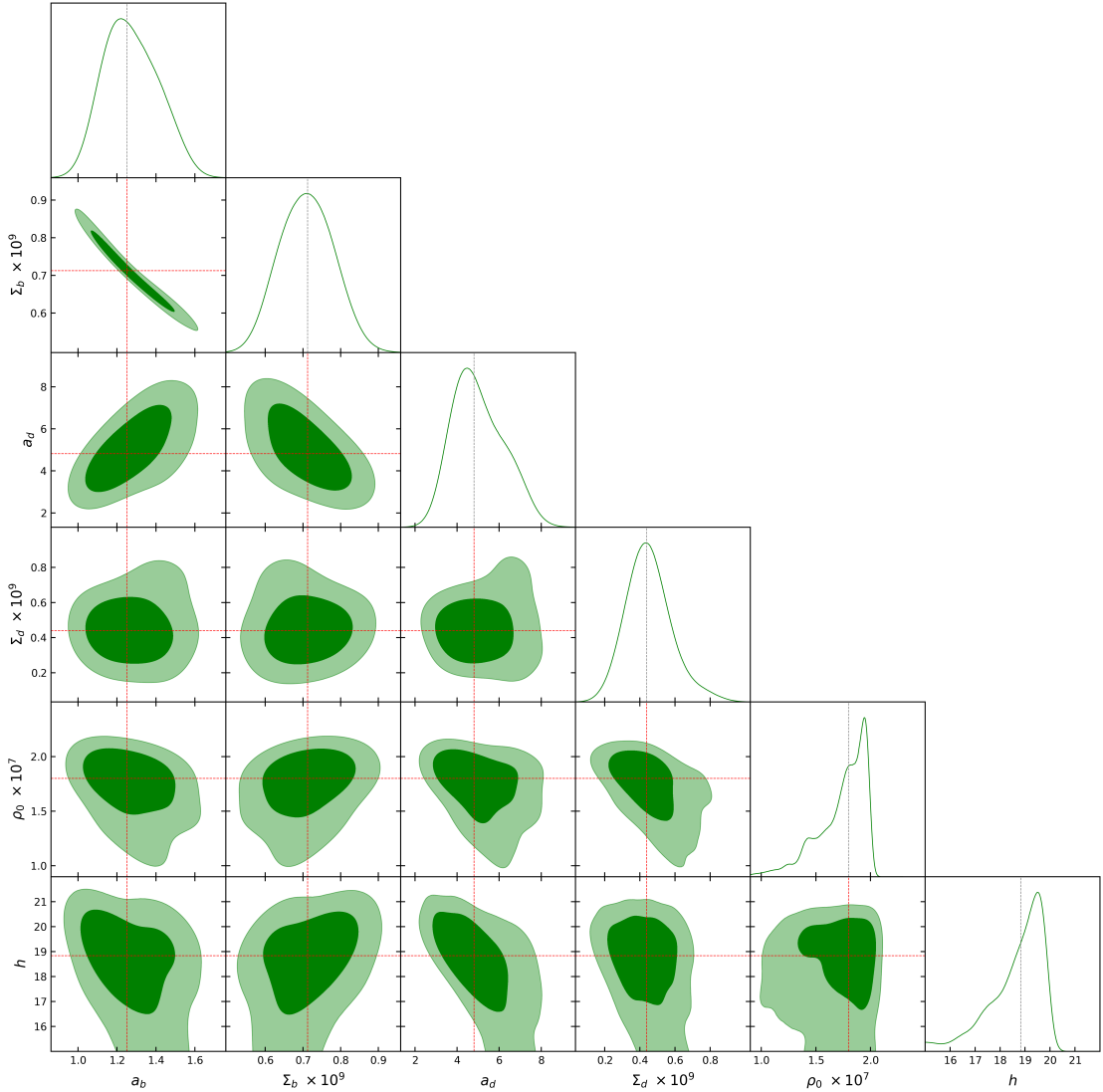

Fig. 5: $1\sigma$ and $2\sigma$ confidence contours and posterior distributions of free parameters of the bulge, disk and halo regions (Hernquist profile) obtained from the MCMC analysis of the M31 galaxy.

4. We first use MCMC method to estimate the parameters of the NFW profile and then used the obtained best fit values to study the rotation curves of the MW galaxy. We find that our results are consistent with the published results (D'Agostino et al. 2024). We extend the analysis on Hernquist and Einasto profiles for both the galaxies (MW and M31). Further, in Figures [7] and [8], we superimpose the reconstructed rotational velocity curves onto the observational data of MW and M31 galaxies, respectively. These reconstructed curves are based on the best-fit values of the bulge, disk, and halo parameters, as listed in Tables 1 and 2.

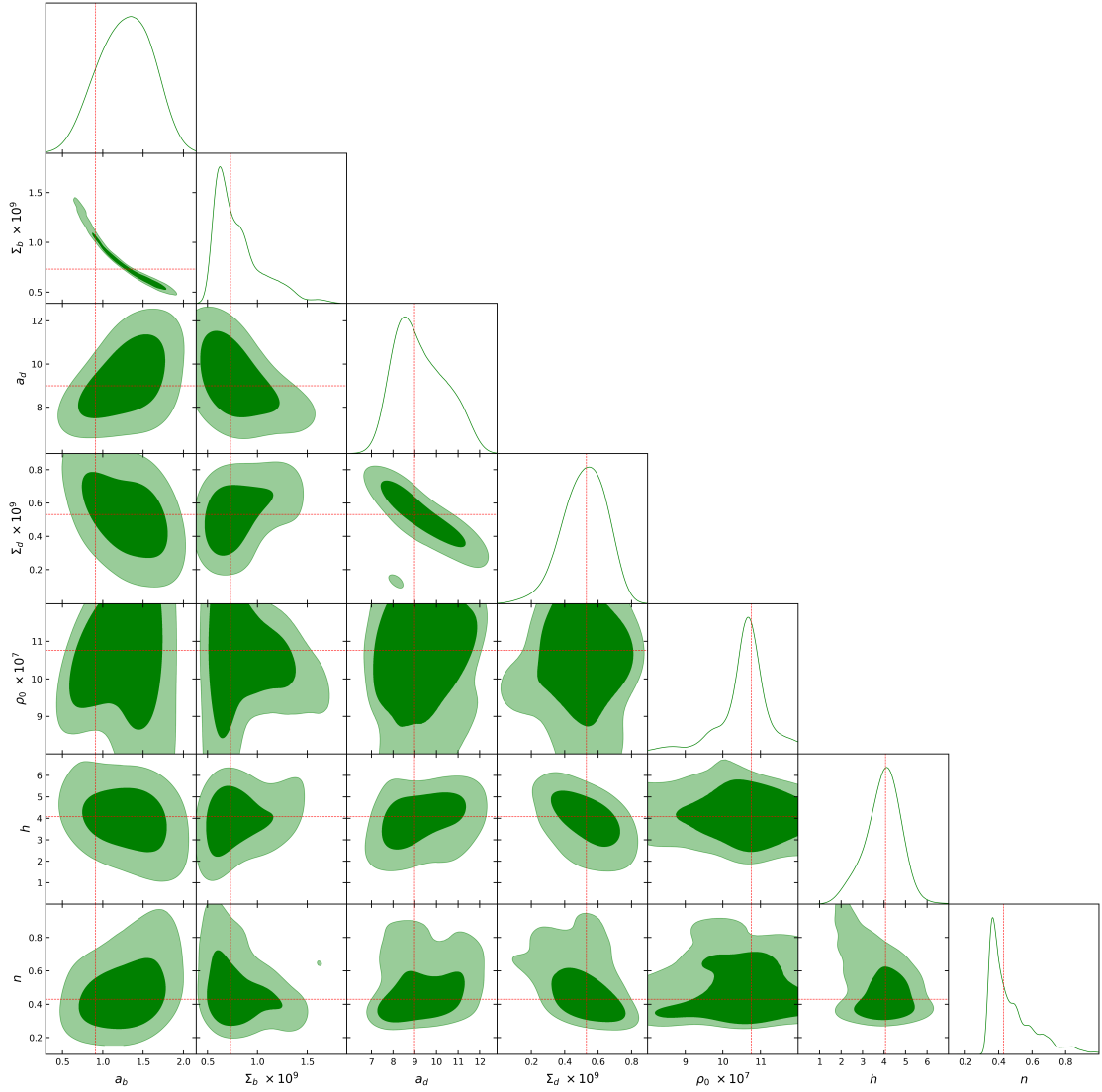

Fig. 6: $1\sigma$ and $2\sigma$ confidence contours and posterior distributions of free parameters of the bulge, disk and halo regions (Einasto profile) obtained from the MCMC analysis of the M31 galaxy.

(a) Our results show that the NFW and Hernquist profiles provide better fits for the dark matter distribution in both MW and M31. The Einasto profile which includes an additional tuning parameter did not perform well, this discrepancy may be due to the limitation of current data and we anticipate that future observations may help to improve constraints on this profile.

(b) We find that the fitted curves for all three profiles in the Milky Way deviate from the observed data points, particularly at low $r$. This discrepancy is primarily due to the fact that the bulge component is not well described by the de Vaucouleurs profile and may consist of multiple subcomponents. For instance, K. Boshkayev et al. considered two subcomponents of the bulge: the inner bulge and the outer bulge and reconstructed the total rotational velocity (Boshkayev et al. 2024). Their fitted curve closely matches the observed data points at smaller $r$. In both the galaxies, we find that the reconstructed curves corresponding to the NFW and Hernquist profiles match the observed data points better than the Einasto profile at high $r$. This variation is likely influenced by the extra parameter included in this profile and the limited number of data points available. L. Chemin also claimed that the Einasto profile declines more strongly than NFW in the outer region, i.e. at higher $r$ (Chemin et al. 2011).

The primary aim of our analysis is to explore various dark matter profiles rather than to carry out an in-depth study of the bulge and disk profiles. When examining the contours of the parameters for the bulge, disk, and halo, we find that the halo parameters are very weakly correlated with those of the bulge and disk. As a result, the best-fit values for the dark matter halo profile parameters are unlikely to be significantly influenced by the choice of different bulge profiles or their subcomponents. This implies that, despite the inaccuracies in fitting the bulge at small $r$, our analysis of various dark halo profiles remains valid, as noted in the work of (Sofue 2015). Therefore, we have not included the different profiles or subcomponents of the bulge in our analysis.

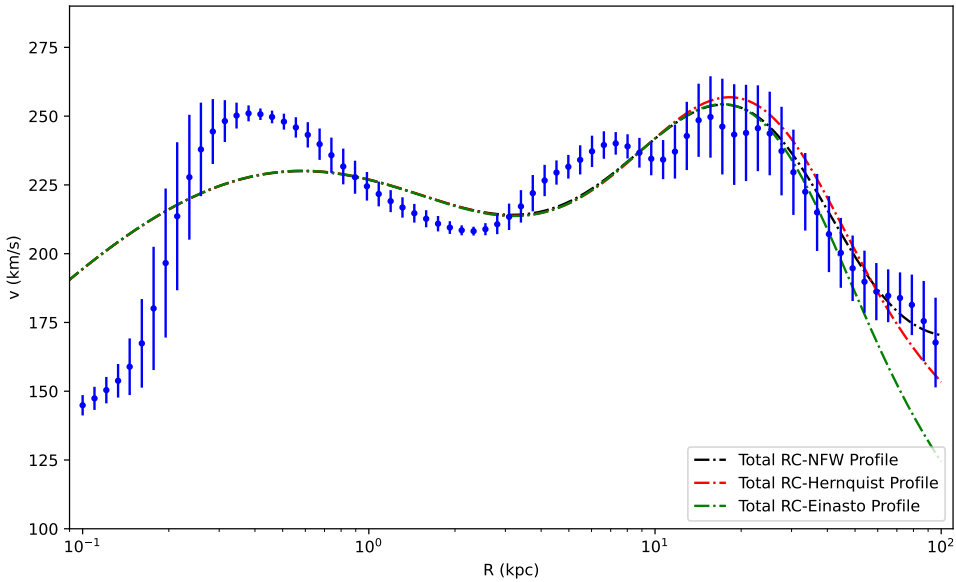

Fig. 7: Reconstructed rotational velocity curves for the NFW, Hernquist, and Einasto profiles, superimposed on the observed data points of the MW galaxy, including error bars.

In conclusion, our work expands on earlier studies by using updated data and introducing the Hernquist and Einasto profiles into the analysis of both MW and M31. While the NFW and Hernquist profiles are robust across both the galaxies, the Einasto profile requires further refinement. Our results provide valuable insights into the dark matter distribution in these two galaxies, contributing to the ongoing effort to understand the structure and evolution of galaxies within the $\Lambda$CDM model. We expect that with the ongoing surveys like Gaia (ESA) [1], Legacy Survey of Space and Time (LSST) [2], Square Kilometer Array (SKA)[3] and future surveys such as Thirty Meter Telescope (TMT)[4] and Extremely Large Telescope (ELT)[5], one will be able to get better constraints on Einasto profile and will also help to resolve the degeneracy among dark halo parameters.

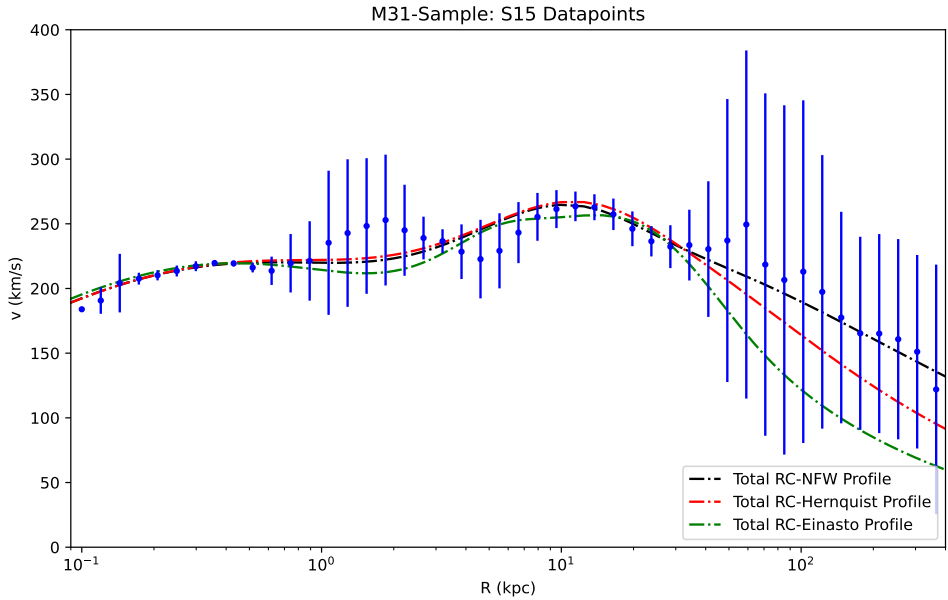

Fig. 8: Reconstructed rotational velocity curves for the NFW, Hernquist, and Einasto profiles, superimposed on the observed data points of the M31 galaxy, including error bars.

## 5 ACKNOWLEDGEMENTS

Darshan is supported by the Startup Research Fund of the Henan Academy of Sciences under grant No. 241841219. N.R. would like to extend her sincere gratitude to the Principal and colleagues of Miranda House, University of Delhi, and to Inter-University center for Astronomy and Astrophysics (IUCAA) Pune.

## References

An, J., & Zhao, H. 2013, MNRAS, 428, 2805  2

Bhattacharjee, P., Chaudhury, S., & Kundu, S. 2014, ApJ, 785, 63  2

Binney, J., & Tremaine, S. 2008, Princeton University Press  3

Boshkayev, K., Konysbayev, T., Kurmanov, Y., et al. 2024, International Journal of Modern Physics D, 33, 2450016  10, 13

Bosma, A. 1981, AJ, 86, 1791  2

Brownstein, J. R. 2009, Modified gravity and the phantom of dark matter, PhD thesis, University of
    Waterloo, Canada 2

Burkert, A. 1995, ApJ, 447, L25 2

Carignan, C., Chemin, L., Huchtmeier, W. K., & Lockman, F. J. 2006, ApJ, 641, L109 2

Chemin, L., de Blok, W. J. G., & Mamon, G. A. 2011, AJ, 142, 109 14

Corbelli, E., Lorenzoni, S., Walterbos, R., Braun, R., & Thilker, D. 2010, A&A, 511, A89 2

Corbelli, E., & Salucci, P. 2000, MNRAS, 311, 441 2

D'Agostino, R., Jusufi, K., & Capozziello, S. 2024, European Physical Journal C, 84, 386 12

Dai, D.-C., Starkman, G., & Stojkovic, D. 2022, Phys. Rev. D, 105, 104067 11

de Vaucouleurs, G. 1948, Annales d'Astrophysique, 11, 247 3

de Vaucouleurs, G. 1959, Handbuch der Physik, 53, 311 4

Einasto, J. 1965, Trudy Astrofizicheskogo Instituta Alma-Ata, 5, 87 5

Foreman-Mackey, D., Hogg, D. W., Lang, D., & Goodman, J. 2013, PASP, 125, 306 5

Freeman, K. C. 1970, ApJ, 160, 811 4

Hernquist, L. 1990, ApJ, 356, 359 5

Huang, Y., Liu, X. W., Yuan, H. B., et al. 2016, MNRAS, 463, 2623 2

Lewis, A. 2019, arXiv e-prints, arXiv:1910.13970 5

Lin, H.-N., & Li, X. 2019, MNRAS, 487, 5679 10

Merritt, D., Graham, A. W., Moore, B., Diemand, J., & Terzić, B. 2006, AJ, 132, 2685 2

Navarro, J. F., Frenk, C. S., & White, S. D. M. 1996, ApJ, 462, 563 2, 4

Navarro, J. F., Frenk, C. S., & White, S. D. M. 1997, ApJ, 490, 493 2

Navarro, J. F., Hayashi, E., Power, C., et al. 2004, MNRAS, 349, 1039 2

Ou, X., Eilers, A.-C., Necib, L., & Frebel, A. 2024, MNRAS, 528, 693 10

Persic, M., & Salucci, P. 1988, MNRAS, 234, 131 2

Persic, M., Salucci, P., & Stel, F. 1996, MNRAS, 281, 27 2

Rubin, V. C., Ford, W. K., J., & Thonnard, N. 1978, ApJ, 225, L107 4

Rubin, V. C., Ford, W. K., J., & Thonnard, N. 1980, ApJ, 238, 471 2

Rubin, V. C., & Ford, W. Kent, J. 1970, ApJ, 159, 379 2

Salucci, P., Esposito, G., Lambiase, G., et al. 2021, Frontiers in Physics, 8, 579 2

Sofue, Y. 2013, PASJ, 65, 118 2

Sofue, Y. 2015, PASJ, 67, 75 2, 3, 5, 14

Sofue, Y. 2017, PASJ, 69, R1 2

Sofue, Y. 2020, Galaxies, 8, 37 3, 5

Zhao, H. 1996, MNRAS, 278, 488 2